\documentstyle[aps,twocolumn]{revtex}
\def\bea{\begin{eqnarray}}
\def\eea{\end{eqnarray}}
\def\be{\begin{equation}}
\def\ee{\end{equation}}
\begin{document}
\author{Krzysztof Sacha$^1$, 
Jakub Zakrzewski$^1$, and Dominique Delande$^2$  }
\address{
 $^1$Instytut Fizyki imienia Mariana Smoluchowskiego, Uniwersytet
Jagiello\'nski,\\
ulica Reymonta 4, PL-30-059 Krak\'ow, Poland\\
$^2$Laboratoire Kastler-Brossel, Tour 12, \'Etage 1, Universit\'e Pierre et
Marie Curie,\\
4 Place Jussieu, 75005 Paris, France
}
\title{Chaotic Rydberg atoms with broken time--reversal symmetry}

\date{\today}

\maketitle

\begin{abstract}
The  dynamics of Rydberg states of atomic hydrogen perturbed
simultaneously by
a static electric field and 
a resonant microwave field of elliptical polarization
is analysed in the quantum perturbative limit of
small amplitudes. 
For some configurations, the secular motion 
(i.e. evolution in time of the elliptical electronic trajectory)
may be classically 
predominantly chaotic.
By changing the
orientation of the static field with respect to the polarization of the
microwave field,
one can modify the global symmetries of the system and break
any generalized time-reversal invariance. This has a dramatic effect
on the statistical properties of the energy levels.
\end{abstract}
\narrowtext

One of the fundamental questions in quantum chaos is the correspondence
between
the classical dynamics of a given Hamiltonian dynamical system and the
statistical properties of the energy spectrum of its quantum counterpart
\cite{bohigas}. Universal fluctuation properties occur when the corresponding 
classical system is either integrable or chaotic.
The spectral fluctuations for a
classically integrable system  obey Poissonian statistics \cite{berry}. 
In another extreme, i.e. ergodicity,
the statistical properties of quantum spectra
can be modeled by ensembles of random matrices, namely with a GOE (Gaussian
orthogonal ensemble) in the case of time-reversal invariant 
(or more generally,
any other antiunitary invariance) systems or with GUE 
(Gaussian unitary ensemble) in the
absence of such a symmetry \cite{bohigas}. For a 
generic Hamiltonian system, where typically chaotic and regular motions
coexist, the 
fluctuations are expected to interpolate between these two limiting
universal cases \cite{brody,izrailev,robnik}. While experimental observations
of transition to chaos have been made for time-reversal invariant systems, 
the system studied below allows for the breaking of the antiunitary symmetry. 
As far as we know, this is the first experimentally realizable example of 
such a situation in a quantum system (for wave chaos experiments see 
\cite{sosto}).

Perturbed Coulomb systems are ideally suited for studies of a manifestation
of classical chaos in quantum mechanics \cite{wk} both theoretically
and {\it experimentally}.
A hydrogen-like atom in an uniform magnetic field 
reveals 
a smooth transition to chaos \cite{wk} with increasing magnetic field.
Similarly the motion of the electron in an atom driven by electromagnetic
radiation undergoes a smooth transition to chaos with increasing field
amplitude \cite{wk}. On the other hand,
the anisotropic Kepler problem (modeling defects in semiconductors),
becomes chaotic as soon as the effective
mass becomes direction dependent \cite{gutz}.  
Different reactions to perturbations are just a manifestation of the
inapplicability of the 
Kolmogorov-Arnold-Moser theorem
\cite{gutz} to a highly degenerate Coulomb problem.

In this letter, we consider the perturbation of the hydrogen (H) 
atom by very weak
static electric and resonant microwave fields. Classically, such fields 
may produce chaotic dynamics in the secular motion of the atom (i.e. in the
motion of the electronic ellipse). 
In the quantum mechanical language, states coming from a given
manifold with
fixed principal quantum number $n_0$
may significantly mix only with each other. Thus, we may observe an
``intramanifold''
chaos, in which a finite number (possibly $n_0^2$) of quantum levels are
involved (provided the coupling to the continuum is negligible). 
Intramanifold chaos requires the strong mixing of at least 2 different degrees
of freedom, here the total angular momentum and its $z$-component.

The simplest system where such a behaviour has been expected is a hydrogen
atom perturbed by uniform  
static electric and magnetic fields of arbitrary
mutual orientation \cite{vonm,main}.
At lowest non-vanishing order (small fields) the perturbation
 leaves a constant of the motion.
By including second order terms, signatures of chaos have been observed in the
quantum spectrum \cite{vonm,main}. The necessary fields are, however,
big and
higher order terms -- neglected in
this approach --
are of comparable importance.
Moreover, the high electric field leads then to
fast field ionization which will blur the effects
discussed.
The situation
discussed below does not suffer from this deficiency. Moreover it allows us
to study the influence of an anti-unitary
symmetry breaking  on the level statistics.

We consider a realistic three-dimensional H atom placed in a static
electric field and driven by elliptically polarized microwaves. We define the
$z$-axis as perpendicular to the plane of polarization of the microwave
field. The Hamiltonian
of the system in atomic units, neglecting relativistic effects, assuming
infinite mass of the nucleus, and employing dipole approximation reads:
\be
H=\frac{\mathbf p^2}{2}-\frac{1}{r}+F(x\cos\omega t+\alpha y\sin\omega
t)+\mathbf
E\cdot\mathbf r,
\label{h}
\ee
where $F$, $\alpha$ and $\omega$ stand, respectively,
for the amplitude, degree of elliptical polarization and frequency of 
microwave field while $\mathbf E$ denotes the static electric field vector.
The resonant driving of Rydberg states with principal quantum number $n_0$
occurs when the atom is illuminated 
by a microwave field of frequency fulfilling $\omega_0=\omega/\omega_K=m$, 
where $m$ is an integer number and $\omega_K=1/n^3_0$ is the frequency of 
the unperturbed Kepler motion. 

Because of time-periodicity, the Hamiltonian has a series of quasi-energy
levels, 
corresponding to the energy levels of the atomic system dressed by the
microwave
photons. To calculate these quasi-energy levels, we employ a quantum
perturbation method,
whose details will be given 
elsewhere. In short, the method defines an effective Hamiltonian
\cite{cohen}
in a given manifold, which takes into 
account the {\it direct} coupling between the levels due to the presence of
the static electric field and the {\it indirect} coupling through 
all levels of other manifolds, i.e. 
process of absorption and emission of microwave photons. The {\it direct} 
couplings gives a first order contribution of the static field while
the {\it indirect} ones contribute at second order of the microwave field.
The effective Hamiltonian inside a given hydrogenic manifold is thus the sum
of a term proportional to $E$ and a term proportional to $F^2.$ For
arbitrary mutual
orientations of the two fields, the two terms have incompatible symmetry
properties
and, when having comparable magnitudes, induce a global chaotic behaviour.
The resulting matrix of dimension $n_0^2$ is diagonalized by standard
routines.
The method has been validated in the limiting case of combined weak static
and linearly polarized microwave fields \cite{ours}.

The semiclassical approach 
is an
extension
of the methods used in our previous studies 
\cite{semicl,abu}. 
It relies on quantization of the fast orbital motion of the electron 
keeping fixed the electronic ellipse. In the resonant case, this motion
is pendulum-like. Taking the appropriate limiting case (valid for
very weak fields) of the corresponding Mathieu equation solution 
\cite{abram} yields the effective 
Hamiltonian for the secular motion of the electronic ellipse, which is the
classical
equivalent of the effective quantum Hamiltonian discussed above. It
expresses as:
\be
H_{\mathrm sec}=\frac{F_0^2}{3m^{2}}\Gamma^2_{m,0}
+\frac{E_0}{n_0^2}\gamma,
\label{final}
\ee
where the constant $-1/2n_0^2$ term was
omitted ($H_{\mathrm sec}$ denotes the shift from
the unperturbed energy level of the atom), and:
\bea
\gamma &=&-\frac{3}{2}\left[\cos\varphi\sin\theta
          \left(\cos\phi\cos\psi-
\frac{M_0}{L_0}\sin\phi\sin\psi\right)\right. \cr
&&  +\sin\varphi\sin\theta\left(\sin\phi\cos\psi+\frac{M_0}{L_0}
\cos\phi\sin\psi\right) \cr
&& \left. +\cos\theta\sqrt{1-\frac{M_0^2}{L_0^2}}\sin\psi
\right]\sqrt{1-L_0^2}.
\label{g}
\eea
$F_0=n_0^4F$, $E_0=n_0^4E$, $L_0=L/n_0$ and $M_0=M/n_0$ are the scaled
microwave
and static electric field amplitudes, the scaled angular momentum and the
projection of the angular momentum on the $z$-axis, respectively. 
The $\psi$ and $\phi$ angles are canonically conjugate variables to the $L_0$
and $M_0$ momenta, respectively.
Orientation of the static field vector with respect to the $z$-axis is
given by the spherical angles $\varphi$ and $\theta$. The expression for
$\gamma$
looks complicated, but it is actually nothing but the component of the 
average atomic dipole on the static field axis. The explicit expression 
for $\Gamma_{m,0}=\Gamma_m/n_0^2$ is given in \cite{expgam}, Eq.~(2.16).

To calculate the quasi-energies of the system, one should perform
quantization 
of the secular motion determined by the Hamiltonian (\ref{final}). 
It has been done in simpler situations (e.g. H atom perturbed by
linearly polarized microwaves and a parallel static electric field) 
 \cite{semicl}. 
Then the secular motion is integrable and its
 quantization straightforward.
The present secular problem (2 degrees of freedom)
 turns out to be non-integrable. 

Eq.~(\ref{final}) has some well defined scaling properties with the
field strengths $F_0$ and $E_0.$ Let us define the reduced microwave strength
${\cal F} = F_0^2n_0^2/E_0 = F^2n_0^6/E$ and the reduced Hamiltonian
\be
{\cal H} =H_{\mathrm sec}n_0^2/E_0=\frac{\cal F}{3m^{2}}\Gamma^2_{m,0}
+\gamma.    
\ee
The classical phase space structure depends only on the value of ${\cal H}$
and
${\cal F}$ (beside the static
field vector orientation and the polarization of the microwave), but not on
the detailed values of $n_0,$ $E,$ $F$ and the secular energy.
Of course, weaker fields imply a slower secular motion, but this does not
affect
the structure of phase space. In the quantum mechanical picture, the
energy splitting of a degenerate hydrogenic manifold also depends 
on absolute values of the fields, but the structure of the levels does not.

For generic orientations of the fields, all symmetry properties are broken. 
It is only for $\varphi$ equal to $0$ or $\pi/2$ 
that the system has some antiunitary invariance, under the combination
of time-reversal 
transformation with reflection with respect to the $YZ$ or 
$XZ$ plane, see (\ref{h}).
To observe the quantum signature of the breaking of the antiunitary symmetry, 
it is convenient to have a predominantely chaotic classical dynamics, as a
transition from GOE to GUE statistics is expected. After rather extensive
searches
with various values of the parameters, we have found that the 2:1 resonance,
i.e. $\omega_0=m=2$, is more suitable than the principal one as
it produces a more chaotic dynamics. The choice $\theta\approx \pi/4,$ 
$\alpha=0.4$ and ${\cal F} = 5000$ maximizes chaoticity of the system for 
reduced energy ${\cal H}$ in the range $8.5-9.5.$ A Poincar\'e surface of
section
for these parameters (and $\varphi=\pi/4$) is shown in Fig.~\ref{one} and
displays
chaos very predominantly. Moreover, the phase space structure 
does not change significantly if we 
set $\varphi$ equal to $0$ or $\pi/2$ (for $\varphi=0$ the regular layer is
slightly larger while for $\varphi=\pi/2$ it disappears). 

To obtain better statistics, the quantum levels were found for different
orientation angles $\varphi,\ \theta=\{0.2\pi,0.25\pi,0.3\pi\}$ and different
$n_0=99-101$ values, 
keeping other parameters as in Fig.~\ref{one}. From 
each diagonalization, we took a fragment of spectrum
corresponding to the quasi-energy interval $8.5-9.5$ and unfolded the spectra
to produce the statistical distributions.

The cumulative NNS distribution is plotted in
Fig.~\ref{two} together with the similar distributions corresponding to GOE
and GUE \cite{bohigas}. It is apparent that the calculated numerical data 
traces between GOE and
GUE, clearly demonstrating the breaking of the antiunitary symmetry.
The data do not reach the GUE behavior because  
the corresponding classical dynamics is not entirely ergodic. 
To measure 
the departure from the GUE distribution, one may fit different theoretical
distributions. A natural choice is Berry-Robnik
statistics \cite{robnik} (for 
independent superposition of Poisson and 
GUE spectra) which allows one to 
estimate the relative measure $q$ of the chaotic part of classical phase
space. 
The best fit is for
$q = 0.94$ which is in agreement with Fig.~\ref{one}. 
However, the agreement with the Berry-Robnik distribution is not
perfect, see Fig.~\ref{two}, which suggests 
that the deep semiclassical limit, required for Berry-Robnik statistics
\cite{robnik}, 
is not reached yet in our data.

The statistical distribution proposed by Izrailev 
\cite{izrailev} is more appropriate. Indeed, it fits the numerical
NNS distribution quite impressively (compare Fig.~\ref{two}) yielding
the ``repulsion parameter''
$\beta=1.47,$ roughly in the middle between the GOE $(\beta=1)$ and
GUE $(\beta=2)$ values. 
 
To focus on the long range correlations in the spectra,
we study also the  spectral rigidity, i.e. $\Delta_3$ statistics
\cite{bohigas}. It gives an independent information about the relative
measure $q$ 
of the chaotic part of phase space. 
For a superposition of independent Poissonian and GUE spectra, one obtains
\cite{bohigas,robnik}:
\be
\Delta_3(L)=\Delta_3^{Poisson}((1-q)L)+\Delta_3^{GUE}(qL).
\label{delta}
\ee
Fitting Eq.~(\ref{delta}) to the numerical spectral rigidity, results in 
$q=0.94$ which is in excellent agreement with the value of $q$
obtained from NNS distribution.
In Fig.~\ref{three}, we also present the spectral rigidities obtained 
for antiunitary symmetry invariant cases, i.e. for $\varphi=0$ and
$\varphi=\pi/2$. 
Their detailed analysis  is left for a future
publication.
It is sufficient to say that, for $\varphi=0,$ the phase space structure is
slightly more regular than one shown in Fig.~\ref{one} and consequently
the fitted parameter $q=0.82$ is smaller [for antiunitary
invariant problem $\Delta_3^{GUE}$ is substituted by $\Delta_3^{GOE}$ in
Eq.~(\ref{delta})]. On the other hand, for $\varphi=\pi/2,$ the classical
phase
space reveals more chaotic character and thus $q=0.95$. Note that the
saturation
of $\Delta_3$, for all
presented cases, appears for $L\approx 40$ \cite{berry}.

We have here considered small but finite field amplitudes to stay well 
within the applicability range of the quantum perturbation theory. We
expect, however, a similar behavior for larger
amplitudes. While we present numerical data for $n_0\approx 100$ for
better statistics, a clear signature is also observed for 
$n_0\approx 50$. There, the expected mean level spacing is of the order
of few MHz (for $F_0\approx 5 \times 10^{-3}$) making the experimental
resolution of
the spectrum feasible.

We are grateful to Felix~Izrailev for the permission to use his code 
for his NNS distribution.
Support of KBN under project 2P302B-00915 (KS and JZ) is acknowledged. 
This work is a part of the Alexander von Humboldt scholarship proposal (KS).
Laboratoire Kastler Brossel de
l'Universit\'e Pierre
et Marie Curie et de l'Ecole Normale Sup\'erieure is
UMR 8552 du CNRS.
The additional support of the bilateral Polonium and PICS programs is
appreciated.

\begin{figure}
\caption{Poincar\'e surface of section (at $\phi=0$) 
of the classical secular motion, Eq.~(\protect{\ref{final}}), 
for ${\cal H}=8.5,$ with parameters of the fields ${\cal F}=5000$,
$\omega_0=2$ (i.e. $m=2$), $\alpha=0.4$. Orientation of the 
static electric field vector 
$\varphi=\theta=\pi/4$. Note that, for the parameters chosen,
not the whole $(L_0,\psi)$ space is accessible.}
\label{one}
\end{figure}

\begin{figure}
\caption{Cumulative level spacing distributions $W(s).$ In panel (a), the
solid line
represents
the numerical data for  $n_0=99-101$,
${\cal F}=5000$, $\omega_0=2$, $\alpha=0.4$ 
and different 
static electric field vector orientations 
$\varphi,\ \theta=\{0.2\pi,0.25\pi,0.3\pi\}$ (there are about 6000 spacings);
dashed line (hardly visible behind the solid line): the best fitting
Izrailev distribution with its parameter $\beta=1.47$;
dotted and dash-dotted lines: GOE and GUE distributions respectively.
Panel (b)
 shows a fine-scale representation of the deviation of the numerical
level spacing distribution from the best fitting Izrailev distribution in 
terms of the $U(W(s))-U(W_{Izrailev}(s))$ vs. $W(s)$; the transformation
$U(W)=\arccos\sqrt{1-W}$ 
is used in order to have uniform statistical error 
over the plot [5].
The upper and lower noisy curves represent one standard deviation from the
calculated numerical data which thus lie in the middle of the band. 
The  solid curve is the best Berry-Robnik distribution, 
$q=0.94$.}
\label{two}
\end{figure}

\begin{figure}
\caption{Spectral rigidity, $\Delta_3$ compared with random ensembles 
predictions. Panel (a) shows the results for numerical data (solid line) in 
the broken anti-unitary invariance case
(with the same parameters as in
Fig.~\protect{\ref{two}}) together with the fit of Eq.~(\protect\ref{delta})
(dots, barely visible behind the solid line, fitted value $q=0.94$). 
Dashed lines indicate Poisson, GOE and
GUE predictions as indicated in the figure. Panel (b) shows numerical data
for an antiunitary invariant case with the same parameters as in
Fig.~\protect{\ref{two}}, but for
 $\varphi=0$ and $n_0=97-102$ for the upper case and
 $\varphi=\pi/2$ for the lower line. The fitted values for the
 fraction of chaotic phase space volume are $q=0.82$ and $q=0.95$,
 respectively.
}
\label{three}
\end{figure}

\end{document}